\documentclass[11pt,a4paper]{article}
\pdfoutput=1
\usepackage{jheppub}
\usepackage{amsfonts,amssymb,epsfig,verbatim,mathbbol,amstext,amsmath,psfrag}
\usepackage{graphicx}
\usepackage[latin2]{inputenc}
\usepackage{t1enc}
\usepackage{amsmath}
\usepackage{subfigure}
\usepackage{dsfont}
\usepackage{slashed}
\usepackage{multirow}
\usepackage{lscape}
\usepackage{wrapfig}

\usepackage{setspace}

\renewcommand{\d}{\textmd{d}}
\newcommand{\be}{\begin{equation}}
\newcommand{\ee}{\end{equation}}
\newcommand{\Tr}{\textmd{Tr}}
\newcommand{\tr}{\textmd{tr}}

\newcommand{\Z}{\mathcal{Z}}
\newcommand{\D}{\mathcal{D}}
\newcommand{\expv}[1]{\left \langle #1 \right \rangle}
\newcommand{\expvB}[1]{\Big \langle #1 \Big \rangle}
\newcommand{\expvb}[1]{\big \langle #1 \big \rangle}
\newcommand{\qt}{q_{\rm top}}
\newcommand{\Qt}{Q_{\rm top}}
\newcommand{\ft}{t}

\newcommand{\THET}{g}

\newcommand{\integ}{}

\newcommand{\si[1]}{\textmd{sign}(#1)}

\newcommand{\qdist}[1]{\langle\!\langle#1\rangle\!\rangle}
\newcommand{\qdistb}[1]{\big \langle\!\big\langle#1\big\rangle\!\big\rangle}

\long\def\symbolfootnote[#1]#2{\begingroup%
\def\thefootnote{\fnsymbol{footnote}}\footnote[#1]{#2}\endgroup} 

\newcommand{\Wuppertal}{Bergische Universit\"at Wuppertal, Theoretical Physics, 42119 Wuppertal, Germany.}
\newcommand{\Budapest}{E\"otv\"os University, Theoretical Physics, P\'azm\'any P. s 1/A, H-1117, Budapest, Hungary.}
\newcommand{\Regensburg}{Institute for Theoretical Physics, Universit\"at Regensburg, D-93040 Regensburg, Germany.}
\newcommand{\Juelich}{J\"ulich Supercomputing Centre, Forschungszentrum J\"ulich, D-52425 J\"ulich, Germany.}
\newcommand{\Tata}{Tata Institute of Fundamental Research, Homi Bhabha Road, Mumbai 400005, India.}
\newcommand{\Lendulet}{MTA-ELTE Lend\"ulet Lattice Gauge Theory Research Group.}

\begin{document}

\title{
Local CP-violation and electric charge separation by magnetic fields from lattice QCD
}

\author[1,2]{G.~S.~Bali,}
\author[1]{F.~Bruckmann,}    
\author[1]{G.~Endr\H{o}di,$^\dagger$}
\author[3,4,5]{Z.~Fodor,}  
\author[4,6]{S.~D.~Katz,}    
\author[1]{A.~Sch\"afer,}    
\affiliation[1]{\Regensburg}
\affiliation[2]{\Tata}
\affiliation[3]{\Wuppertal}
\affiliation[4]{\Budapest}  
\affiliation[5]{\Juelich}
\affiliation[6]{\Lendulet}

\emailAdd{gergely.endrodi@physik.uni-regensburg.de}

\abstract{
We study local CP-violation on the lattice 
by measuring the local correlation between the topological charge 
density and the electric dipole moment of quarks, induced by a constant external magnetic field. This 
correlator is found to increase linearly with the external field, with the coefficient of 
proportionality depending only weakly on temperature.
Results are obtained on lattices with various spacings, and are 
extrapolated to the continuum limit after the renormalization of the observables is carried out.
This renormalization utilizes the gradient flow for the quark and gluon fields. 
Our findings suggest that the strength of local CP-violation in QCD with physical quark masses 
is about an order of magnitude smaller than a model 
prediction based on nearly massless quarks 
in domains of constant gluon backgrounds with topological charge.
We also show numerical evidence that the observed local CP-violation 
correlates with spatially extended electric dipole structures in the QCD vacuum.
}

\keywords{chiral magnetic effect, finite temperature, lattice QCD, external field}

\maketitle

\section{Introduction}

Quantum Chromodynamics (QCD) is the theory of the strong interactions. At low temperatures QCD is confining, implying 
that the elementary particles of the theory - quarks and gluons - only exist as components of bound states (hadrons). 
The asymptotic freedom property of QCD ensures that at high temperatures the interaction 
between quarks and gluons weakens, and a transition to the quark-gluon plasma (QGP) phase occurs, where the dominant 
degrees of freedom are no longer colorless bound states but colored objects. According to lattice simulations, this 
transition is no real phase transition but an analytic crossover~\cite{Aoki:2006we} and takes place at around 
$T_c\sim150 \textmd{ MeV}$, see e.g. Refs.~\cite{Aoki:2006br,Borsanyi:2010bp}.

The high-temperature QGP phase is routinely produced in contemporary high energy heavy-ion 
collisions, for example at the Relativistic Heavy Ion Collider (RHIC), where temperatures
exceeding $T_c$ can be reached~\cite{Adare:2009qk}. 
Besides extreme temperatures, another interesting feature of such a heavy-ion collision is 
the presence of strong magnetic fields generated by the spectator particles in non-central events. 
This magnetic field is perpendicular to the reaction plane and may reach values up to $\sqrt{eB}\sim0.1$ 
GeV for RHIC and $\sqrt{eB}\sim 0.5$ GeV for the Large Hadron Collider (LHC)~\cite{Skokov:2009qp}, 
depending on the beam energy and centrality. 
Even though the generated magnetic field has a very short lifetime, of the order of $1$ fm/$c$, this magnetic `pulse' coincides with the formation of the quark-gluon plasma and thus may play an important role in the description of the collision. Strong 
magnetic fields also represent an important concept for cosmology~\cite{Vachaspati:1991nm} and for the description of dense neutron 
stars called magnetars~\cite{Duncan:1992hi}. Therefore, a clear theoretical understanding of the response of QCD matter to external 
magnetic fields is desirable. 

An important characteristic of the QCD vacuum is its transformation property under parity (P) and 
charge 
conjugation (C). In the absence of a $\theta$-parameter, the theory prohibits 
violation of both the P- and CP-symmetries. Indeed, experimental bounds -- 
mostly coming from measurements of the electric dipole moment of the neutron -- 
on the degree of this violation turn out to be extremely tiny. 
Nevertheless, CP-violation could still be realized in the local sense, through fluctuations of CP-odd observables.  
One manifestation of this in the QGP phase created in heavy-ion collisions might be 
through the presence of domains with a non-trivial topological structure of the gluon fields (see, e.g., Ref.~\cite{Kharzeev:1998kz}). 
Such a nonzero topology is indicated by 
the non-vanishing value of the topological charge $\Qt$ (defined below) within that particular domain. 
Since the magnetic field is odd under CP transformation, it is natural to expect that it can be used to effectively 
probe the CP-odd domains of the quark-gluon plasma and, thus, the CP-violating fluctuations in the QCD vacuum.

A possible realization of the coupling between the strong magnetic field and the non-trivial topological structure of the QGP 
is the so-called chiral magnetic effect (CME)~\cite{Kharzeev:2007jp,Fukushima:2008xe}. For close to massless quarks, 
helicity is an approximately conserved quantity, and in strong magnetic fields the quark spins tend to align themselves either parallel 
(for positive charges) or antiparallel (for negative charges) to the external field. Therefore, right-handed, positively 
charged quarks and left handed, negatively charged quarks will tend to have their momenta parallel to the 
direction of the magnetic field. In a domain of the quark-gluon plasma with nonzero topological charge density, there is an 
imbalance between the number of left- and right-handed quarks, due to the Atiyah-Singer index theorem. 
As a consequence, a net 
current of quarks can be produced (anti)parallel to the external magnetic field, or, equivalently, the domain in question will 
be electrically polarized in the direction of the magnetic field.
An alternative formulation of the effect is in terms of a chiral 
chemical potential~\cite{Fukushima:2008xe}, 
which couples to the anomalous axial current and creates a chiral imbalance 
by preferring right-handed over left-handed quarks.

The effects of the electric polarization of the plasma domains 
may persist at later stages of the collision. After hadronization takes place, this 
can result in a preferential emission of charged particles above and below the reaction 
plane~\cite{Kharzeev:2004ey,Voloshin:2004vk}. 
Indications for such a charge asymmetry were observed in the STAR experiment at 
RHIC~\cite{Voloshin:2008jx,Abelev:2009ac} and in the ALICE experiment at the LHC~\cite{Selyuzhenkov:2011xq}. However, to access observables 
related to the CME, 
certain parity-even experimental backgrounds have to be taken into account, which 
complicates the interpretation of the observed data. 
Thus, the exact meaning of these results is still debated, see, e.g., Refs.~\cite{Wang:2009kd,Muller:2010jd,Voronyuk:2011jd,Bzdak:2012ia}.
For recent reviews on the subject see, e.g., Refs.~\cite{Fukushima:2012vr,Kharzeev:2013ffa}.

\begin{figure}[ht!]
\centering
\includegraphics*[width=12cm]{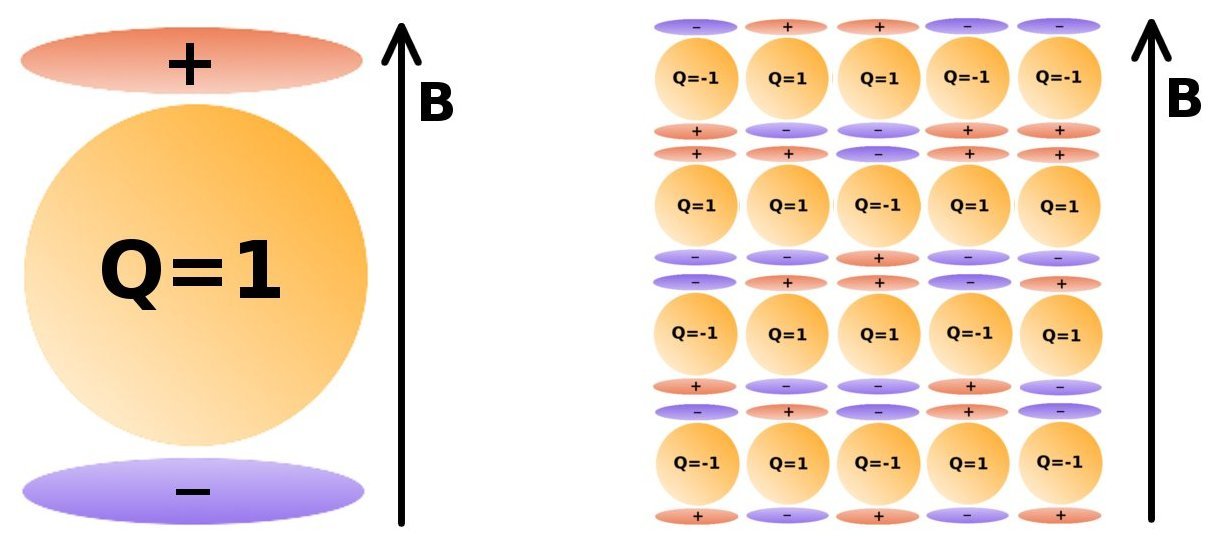}
\caption{\label{fig:illustr}Illustration of the chiral magnetic effect on a single instanton configuration (left panel) and in the QCD vacuum through local fluctuations of the topological charge (right panel).
}
\end{figure}

The CME and topology-induced CP-violation have been studied in various approaches, ranging from effective theory/model calculations 
to Euclidean lattice simulations. 
The former include among others settings like the Nambu-Jona-Lasinio model with an additional coupling to the Polyakov loop 
(PNJL model)~\cite{Fukushima:2010fe}, the holographic approach~\cite{Yee:2009vw,Rebhan:2009vc}, 
hydrodynamics (see, e.g., Refs.~\cite{Zakharov:2012vv,Kalaydzhyan:2012ut}) or using a chiral effective action~\cite{Khaidukov:2013sja}.
On the lattice, the CME was first studied by measuring current- and chirality fluctuations 
in quenched $\mathrm{SU}(2)$~\cite{Buividovich:2009wi} and quenched $\mathrm{SU}(3)$ 
gauge theory~\cite{Braguta:2010ej}. Surprisingly, around the transition temperature, the 
fluctuations of the current parallel to the magnetic field were found to decrease with 
growing $B$ in the small magnetic field region~\cite{Buividovich:2009wi}, a result which still lacks a qualitative 
understanding.  
Another approach to investigate the CME on the lattice is using the 
chiral chemical potential, see Refs.~\cite{Yamamoto:2011gk,Buividovich:2013hza}. 
Finally, the interplay between magnetic fields and topology
was also studied by discretizing a continuum instanton configuration, and 
measuring the electric polarization in the presence of the magnetic field~\cite{Abramczyk:2009gb}, 
see the illustration in the left panel of Fig.~\ref{fig:illustr}.

In the present paper, we pursue a different approach and measure the extent to which the 
topological charge and the electric polarization of the quarks correlate {\it locally}, when 
exposed to external magnetic fields. 
Instead of having to consider classical instanton configurations, this approach enables us to use 
real QCD gauge backgrounds and to consider 
the local fluctuations of the topological charge on them, see the right panel of Fig.~\ref{fig:illustr}. 
Moreover, while 
there is no need to introduce any anomalous current or chemical potential, the method still gives 
a handle on relating the topological and the electromagnetic properties of the QCD vacuum in a 
Lorentz invariant manner.
This approach is similar to that of Ref.~\cite{Buividovich:2009my}, where chirality--electric 
polarization correlators were measured in quenched $\mathrm{SU}(2)$ gauge theory to 
detect the induced electric dipole moment of valence quarks. However, 
in our case the quarks and the external magnetic field are introduced dynamically, 
which allows us to observe spatially extended electric dipole structures 
in the QCD vacuum.

We indeed find that in local domains with nonzero topological charge density, an electric dipole 
moment is induced parallel to the external field. 
The strength of this effect is determined for various magnetic fields and temperatures around 
$T_c$ for several different lattice spacings. 
A scheme for defining the continuum limit of the results, utilizing the 
gradient flow~\cite{Luscher:2010iy} of the fields, is also introduced 
and used to perform the continuum extrapolation.
Finally we compare the lattice results to a model calculation 
that employs nearly massless quarks and constant (anti)selfdual 
gluon backgrounds -- a setting in which the problem can 
be treated analytically~\cite{Basar:2011by}. 
This comparison reveals that the numerical result found for full 
non-perturbative QCD with physical quark masses is 
by an order of magnitude smaller than the model prediction. 

\section{Formulation}

In our setup, we consider the {\it local} correlation of the quark electric dipole moment with the topological charge density,
\be
\qt(x) = \frac{1}{32\pi^2}\, \epsilon_{\mu\nu\alpha\beta}\, \tr\, G_{\mu\nu} (x) \,G_{\alpha\beta} (x),
\label{eq:qtopdef}
\ee
where $G_{\mu\nu}(x)$ is the $\mathrm{SU}(3)$ field strength at the point $x$, 
and $\tr$ denotes the trace in color space. The space-time integral of $\qt$ 
gives the total topological charge $\Qt$. 
In order to define the local electric dipole moment operator, 
let us consider the spin polarization of the 
quark of flavor $f$ (represented by the field $\psi_f$),
\be
\Sigma^f_{\mu\nu}(x) \equiv \bar\psi_f \sigma_{\mu\nu} \psi_f (x), \quad\quad \sigma_{\mu\nu} = \frac{1}{2i} [\gamma_\mu,\gamma_\nu],
\label{eq:smndef}
\ee
where $\gamma_\mu$ are the Euclidean Dirac matrices. 
In the presence of a constant Abelian external field $F_{\mu\nu}$, the spin polarization develops a 
nonzero expectation value~\cite{Ioffe:1983ju},
\be
\expvB{ \integ \Sigma^f_{\mu\nu} (x) }
=q_f F_{\mu\nu} \cdot \tau_f,
\label{eq:chiFf}
\ee
where $q_f$ is the charge of the quark of flavor $f$, and the factor of proportionality $\tau_f$ is conventionally 
written as the product $\tau_f=\expv{\bar\psi_f\psi_f} \chi_f$ of the quark condensate and the 
magnetic susceptibility. Note that the expectation value in Eq.~(\ref{eq:chiFf}) 
involves an integral over space-time and a normalization by the four-volume, to exploit translational invariance.

The $xy$ component of Eq.~(\ref{eq:chiFf}) is induced by an external magnetic field $F_{xy}=B_z$,
whereas the $zt$ component by an external (Euclidean) electric field $F_{zt}=E_z$.
Accordingly, the polarizations correspond to a magnetic 
and an electric dipole moment of the quark, respectively\footnote{Note that this definition of the electric 
dipole moment is normalized with respect to the quark charge $q_f$. To compare to, e.g. Ref.~\cite{Kharzeev:2007jp}, 
one should consider $q_f\cdot \Sigma^f_{zt} $. 
}. 
The electric dipole moment is a parity-odd quantity, just as 
the topological charge density of Eq.~(\ref{eq:qtopdef}); 
their product is therefore parity-even, and can have a nonzero 
expectation value in the presence of the external magnetic field. 
We consider this product locally and write it,  
similarly to Eq.~(\ref{eq:chiFf}), as
\be
\frac
{\displaystyle{}
\expvB{\integ \qt (x)\cdot \Sigma^f_{zt} (x) }}
{\displaystyle{}{
\sqrt{\expvB{\integ \qt^2(x)}}}
}
= q_f B_z \cdot \hat{\tau}_f ,
\label{eq:hatchiFf}
\ee
where we factored out the magnitude of the topological fluctuations 
to define the correlator of the two quantities. 
A similar combination was studied in Ref.~\cite{Buividovich:2009my}. 
Since $\expv{\qt}=0$ we use the square root of the expectation value 
of $\qt^2$ for the normalization. In this way, similarly to 
Eq.~(\ref{eq:chiFf}), we obtain an observable with mass dimension 3, 
and introduce the proportionality factor $\hat{\tau}_f$. 
We emphasize that we consider the product of the two densities on the 
left hand side locally, in order to see the local 
correlation of topology and the polarization (this local product contains 
contact terms which need to be removed by an adequate renormalization 
prescription, see Sec.~\ref{sec:obsrenorm} below). 
Eq.~(\ref{eq:hatchiFf}) expresses the fact that 
there is a local correlation between the topological charge 
density of the non-Abelian vacuum and the induced electric dipole moment, 
and that this correlation is proportional to the external magnetic field\footnote{
This mechanism may be compared to the Witten-effect, through which a magnetic monopole 
develops an electric charge via interacting with a (CP-odd) axion field~\cite{Witten:1979ey}.}.

Consider now the ratio of Eq.~(\ref{eq:hatchiFf}) and the $xy$ component of Eq.~(\ref{eq:chiFf}). 
Here the external field cancels to leading order, giving 
directly the ratio $\hat{\tau}_f/\tau_f$,
\be
C_f \equiv \frac{\hat\tau_f}{\tau_f} = 
\frac
{\displaystyle{}
\expvB{ \integ \qt(x) \cdot \Sigma^f_{zt}(x)}}
{\displaystyle{}
{\sqrt{\expvB{\integ \qt^2(x)}}}
\expvB{\integ\Sigma^f_{xy}(x)}},
\label{eq:Cdef}
\ee
which has dimension zero, and is particularly suited for the lattice determination. 
Note that in this ratio all multiplicative renormalization factors cancel.

\section{Observables and renormalization}
\label{sec:obsrenorm}

We calculate the expectation values appearing in Eq.~(\ref{eq:Cdef}) on the lattice with an 
external magnetic field in the positive $z$ direction, $B_z\equiv B$. The lattice geometry is 
$N_s^3\times N_t$, and the lattice spacing is denoted by $a$, such that 
the spatial volume of the system is given by $V\equiv(aN_s)^3$ and the 
temperature by $T=(aN_t)^{-1}$.
We consider the three lightest quark flavors $u,d$ and $s$, for which the charges are 
$q_u/2=-q_d=-q_s=e/3$ 
(here $e>0$ is the elementary charge). 
We derive our observables from the QCD partition function, which, in the staggered discretization 
of the fermionic action reads
\be
\Z = \int \D U e^{-\beta S_g} \prod_{f=u,d,s} \det M_f^{1/4},
\label{eq:partfunc}
\ee
where $\beta=6/g^2$ is the inverse gauge coupling, $S_g$ the gauge action and 
$M_f=M_f(U,q_fB,m_f)=\slashed{D}(U,q_fB)+m_f\mathds{1}$ the fermion matrix, for which we apply two steps of stout smearing on the gluonic links $U$. 
The quark masses are tuned along the line of constant physics (LCP) 
as $m_u=m_d<m_s$, ensuring that the 
isospin averaged zero-temperature hadron masses equal their experimental values~\cite{Borsanyi:2010cj} 
(for the present action the most precise LCP can be read off from Fig.\! 1 
of Ref.~\cite{Borsanyi:2013bia}). 
Further details of the action and the simulation setup can be found in 
Refs.~\cite{Aoki:2005vt,Borsanyi:2010cj,Bali:2011qj}. 
Since the external field couples directly only to quarks, $B$ enters only through the fermion 
determinants. Note that the dependence on $B$ is always of the form 
of the renormalization group invariant combination $q_fB\sim eB$.

For the gauge action $S_g$, we use the tree-level improved Symanzik action, which contains the product of links 
along closed loops of size $1\times1$ (the plaquettes $P_{\mu\nu}$) and of size $2\times1$. The topological 
charge~(\ref{eq:qtopdef}) at the space-time point $x$ 
can be calculated via the field strength $G_{\mu\nu}(x)$, which 
can be discretized as the sum of the antihermitian part of the four 
plaquettes touching the site $x$,
\be
G_{\mu\nu}(x) = \frac{1}{2}\left[W_{\mu\nu}(x)-W_{\mu\nu}^\dagger(x)\right],\quad\quad W_{\mu\nu}(x) = \frac{1}{4} \sum_{x\in P_{\mu\nu}} P_{\mu\nu},
\label{eq:measFS}
\ee
and the product in the four plaquettes starts at point $x$ and advances counter-clockwise. 
To suppress the noise originating from short-range fluctuations, the links used in Eq.~(\ref{eq:measFS}) are the twice 
stout smeared links that we also use in fermionic observables. 
We find that this choice for the definition of $G_{\mu\nu}$ -- and, 
thus, of $\qt$ -- reduces the noise in the correlation 
between $\qt$ and the electric dipole moment, 
necessary for the coefficient $C_f$ of Eq.~(\ref{eq:Cdef}). 
Note that the continuum limit of $C_f$ is unaffected by this choice. 
Let us add here that it is customary to use improved definitions of $\qt$ (see, e.g., 
Ref.~\cite{DeGrand:1997gu}) or much more extensive smearing 
of the gluonic links in order to obtain an integer value for the total 
topological charge $\Qt$. Here we do not aim to determine the total charge, or 
its susceptibility, but concentrate on {\it local} fluctuations in $\qt$ and 
its correlation with fluctuations of the electric dipole moment, 
for which we carefully checked that our setup is appropriate.

The expectation value of the spin polarization with respect to the partition 
function~(\ref{eq:partfunc}) reads
\be
\expvB{\integ\Sigma^f_{\mu\nu}(x)} = \frac{T}{V}\cdot \frac{1}{4} \expv{\Tr(\sigma_{\mu\nu}M_f^{-1})},
\label{eq:pbpTdef}
\ee
where the trace (in color and coordinate space) is determined using noisy estimators $\eta^i$,
such that the polarization at point $x$ is (color indices are suppressed here)
\be
\Sigma^f_{\mu\nu}(x) \approx \frac{1}{N_v}\sum_{i=1}^{N_v} \sum_{y} \,\eta^{i\dagger}_x\, [\sigma_{\mu\nu}M_f^{-1}]_{xy}\, \eta^i_y,
\label{eq:measpol}
\ee
with no summation over $x$. Here, $N_v$ is the number of estimators, which we set, depending on the ensemble, in the range $40\ldots80$. Furthermore, $\sigma_{\mu\nu}$ stands 
for the staggered representation of the tensor operator, see Ref.~\cite{Bali:2012jv} 
for the implementation we use.

Using the expressions~(\ref{eq:qtopdef}),~(\ref{eq:measFS}),~(\ref{eq:pbpTdef}) and~(\ref{eq:measpol}), the expectation values appearing in the 
ratio $C_f$ 
of Eq.~(\ref{eq:Cdef}) are determined. The so obtained $C_f$ is yet to be renormalized, since both its numerator and 
denominator contain divergent terms. The magnetic dipole 
moment, for example, contains a logarithmic additive divergence, 
which may be eliminated using the operator $m_f\partial/\partial m_f$, see 
Ref.~\cite{Bali:2012jv}. The square of $\qt$ is also subject to renormalization, as it contains the 
contact term, see, e.g., Ref.~\cite{Bruckmann:2011ve}. 
Similarly, one expects the numerator to contain 
terms that are infinite in the continuum limit. These divergences are related to the fact that two densities are 
multiplied at the same space-time point. 
To remove these unphysical contributions, we use the gradient 
flow~\cite{Luscher:2010iy} for the 
fields contained in $\qt$ and in $\Sigma^f_{\mu\nu}$. 
The gradient flow was shown to eliminate additive divergences in fermionic observables like the condensate or the pseudoscalar 
correlator~\cite{Luscher:2013cpa}.
Likewise, we find that evolving the fields up to a fixed physical flow time 
$\ft^{\rm ph}$ -- or, 
equivalently, applying a nonzero smearing range $R_s=\sqrt{8\,\ft^{\rm ph}}$ -- 
renormalizes the observable $C_f$ and, 
at the same time, suppresses noise considerably. 
Our implementation of the gradient flow is detailed in App.~\ref{app:flow}.

Finally, the operator $\Sigma^f_{\mu\nu}$ is also subject to multiplicative renormalization by the tensor 
renormalization constant $Z_{\rm T}$, which was calculated in perturbation theory for the present action 
in Ref.~\cite{Bali:2012jv}. However,
this factor cancels in the ratio $C_f$. 
Altogether, $C_f$ is ultraviolet finite, if the continuum limit is approached along a fixed nonzero smearing range $R_s$. On the lattice, 
this corresponds to taking the limit $N_t\to\infty$ at a fixed temperature $T=(N_ta)^{-1}$ and tuning the smearing range in lattice units as $R_s^{\rm lat} = R_s / a$. 
We repeat the continuum extrapolation for several ranges $R_s>0$ and subsequently extrapolate the results to $R_s=0$. 

Let us point out that in the present study smearing is applied in two different 
contexts. First, stout link smearing is employed in the fermionic action in 
order to suppress lattice discretization errors and, thus, to improve the convergence 
towards the continuum limit. 
Second, the fields in certain observables 
are evolved according to the gradient flow, which is equivalent 
to performing infinitesimal smearing steps. 
The latter reduces unphysical ultraviolet 
contributions in some observables, allowing for a clean definition 
of the continuum limit. 

\section{Results}
\label{sec:results}

\begin{figure}[ht!]
\centering
\includegraphics*[width=8cm]{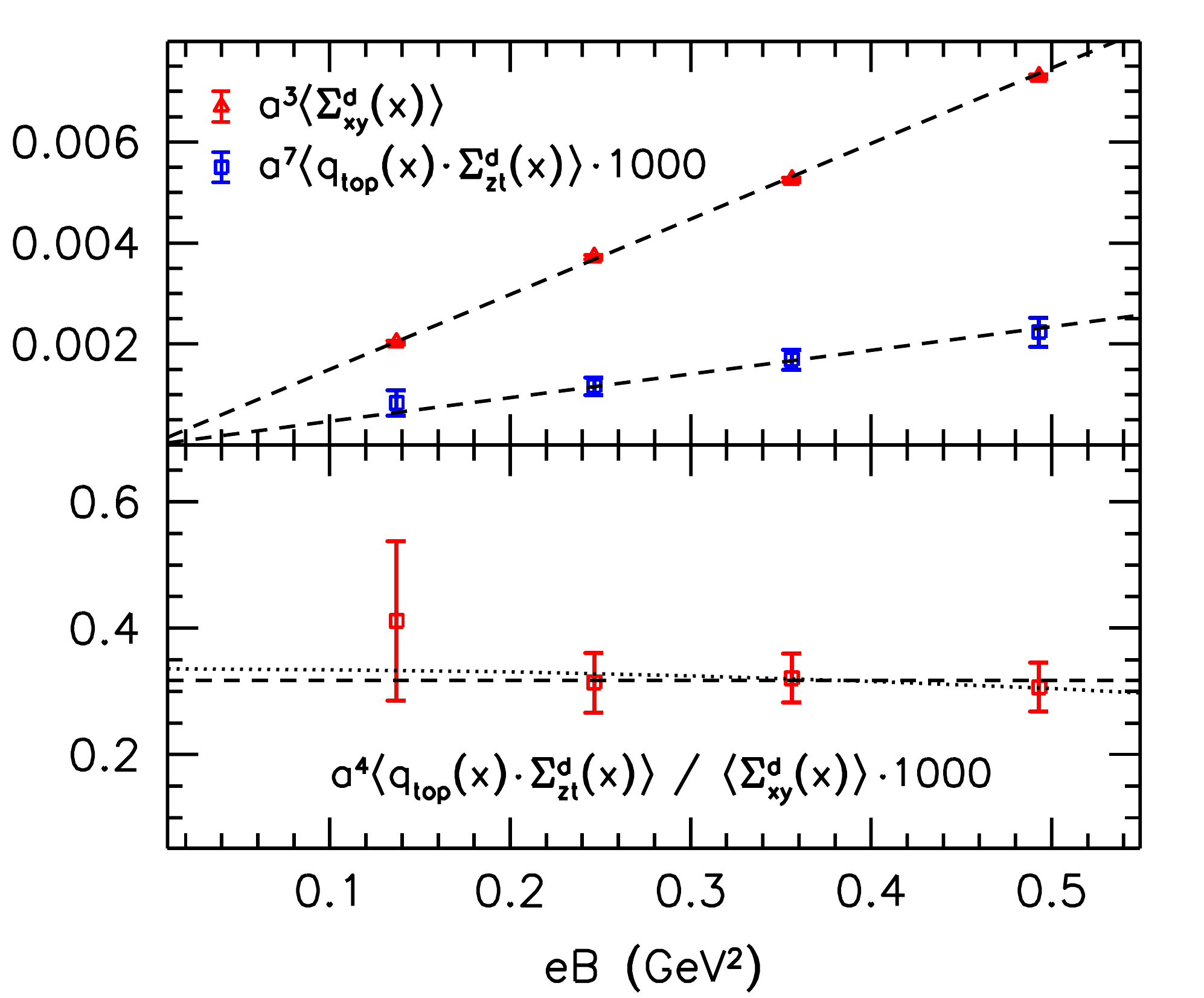} \quad
\includegraphics*[width=8.4cm]{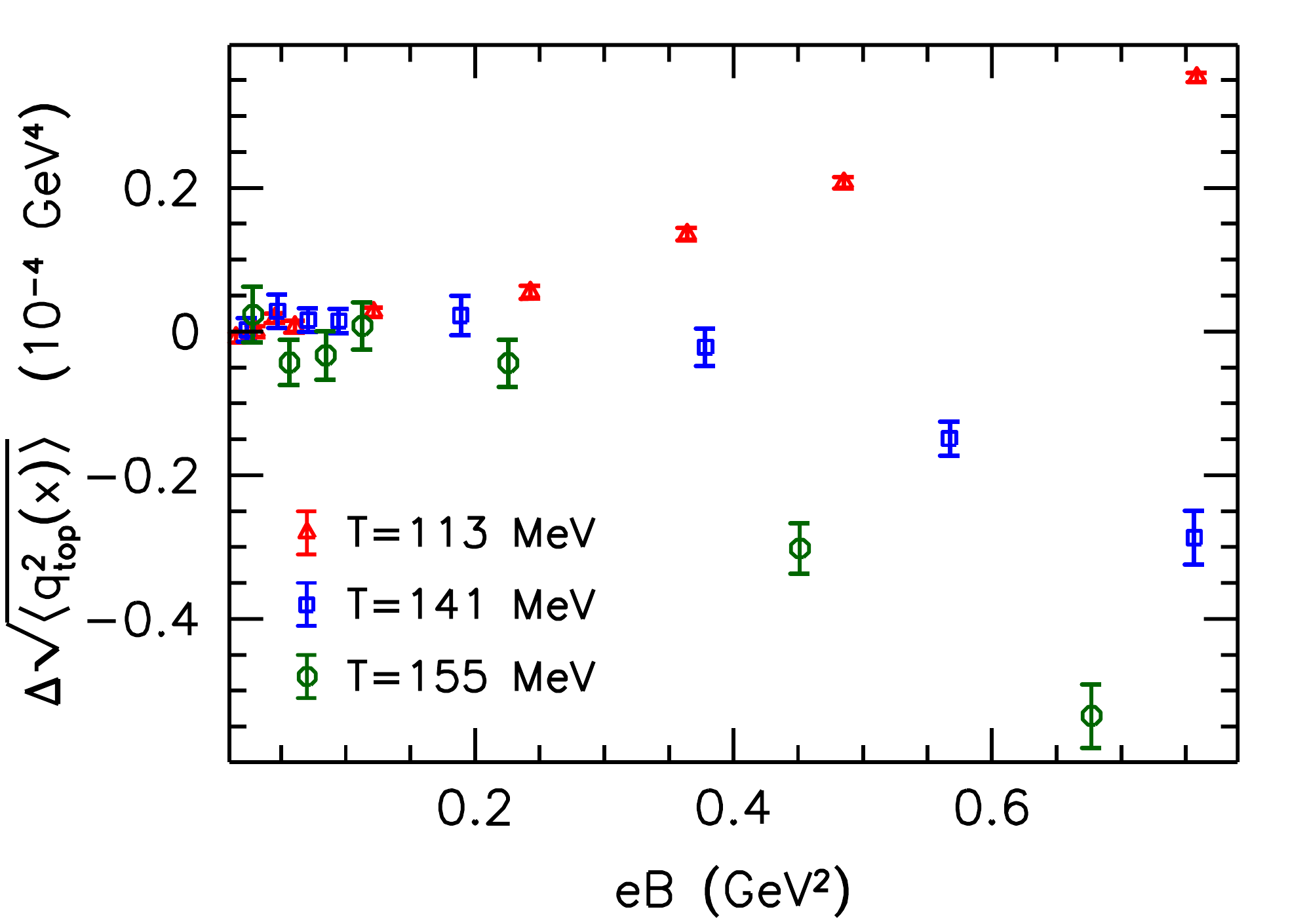}
\caption{Upper left panel: the magnetic dipole moment (red triangles) and the 
correlator of the topological charge density with the 
electric dipole moment (blue squares) in lattice units, with linear fits.
Lower left panel: the ratio of the above two quantities, with a constant (dashed line) 
and a quadratic fit (dotted line). 
The data correspond to a temperature $T=113 \textmd{ MeV}$, as measured on 
the $24^3\times 8$ lattice ensemble. 
Right panel: change in the local fluctuations of the topological charge density 
due to the magnetic field for a few temperatures 
below and around the transition region, as measured on the $24^3\times 6$ ensemble.
}
\label{fig:dipmoms}
\end{figure}

We first analyze the response of $\expvb{\qt(x) \cdot \Sigma^f_{zt}(x)}$ to the external magnetic field. 
Together with the results for $\expvb{\Sigma^f_{xy}(x)}$, this is plotted for the down quark in the upper left panel 
of Fig.~\ref{fig:dipmoms}. The ratio of the two expectation values is expected to be independent of the magnetic field, up to 
corrections of $\mathcal{O}((q_fB)^2$), in accordance with Lorentz invariance. Within the range of the applied magnetic fields, these corrections are found to be small, and thus the ratio is to a good approximation constant, see the lower left panel of Fig.~\ref{fig:dipmoms}. In order to determine the leading order $B$-dependence of the ratio, in the following we fit the data 
either to a constant, or consider corrections of $\mathcal{O}((q_fB)^2)$. Our strategy for the determination of the systematic error of the 
result will be discussed below.

The next step to obtain the coefficient $C_f$ of Eq.~(\ref{eq:Cdef}) is to measure 
the local fluctuations\footnote{Note 
that $\expvb{\qt^2(x)}$ measures the extent of local fluctuations, 
in contrast to the topological susceptibility $\expvb{\Qt^2}\sim \chi_{\rm top}$, 
which quantifies the global fluctuations.} in $\qt$. We 
find that $\expvb{\qt^2(x)}$ depends quadratically on $eB$ 
(again in accordance with Lorentz invariance), however, with a 
coefficient that changes sign as the temperature is 
increased across the transition temperature $T_c$. This behavior is reminiscent of that of the chiral 
condensate~\cite{Bali:2011qj,Bali:2012zg,Bruckmann:2013oba} 
as well as of the gluonic action~\cite{Bali:2013esa}, which 
undergo magnetic catalysis at low temperatures and inverse catalysis in the transition region. The 
change in the local fluctuations due to the magnetic field,
$\Delta \sqrt{\expv{\qt^2(x)}} = \left.\sqrt{\expv{\qt^2(x)}}\right|_B - \left.\sqrt{\expv{\qt^2(x)}}\right|_0$
 is shown in the right panel of Fig.~\ref{fig:dipmoms} for different temperatures.
We note that although this change is significant, its magnitude is negligible compared to 
$\sqrt{\expvb{\qt^2(x)}}$ at $B=0$ for the magnetic fields under study, in accordance with the findings for the two-point function 
of the topological charge density in Ref.~\cite{Bali:2013esa}. 

\begin{figure}[ht!]
\centering
\mbox{
\includegraphics*[width=8.5cm]{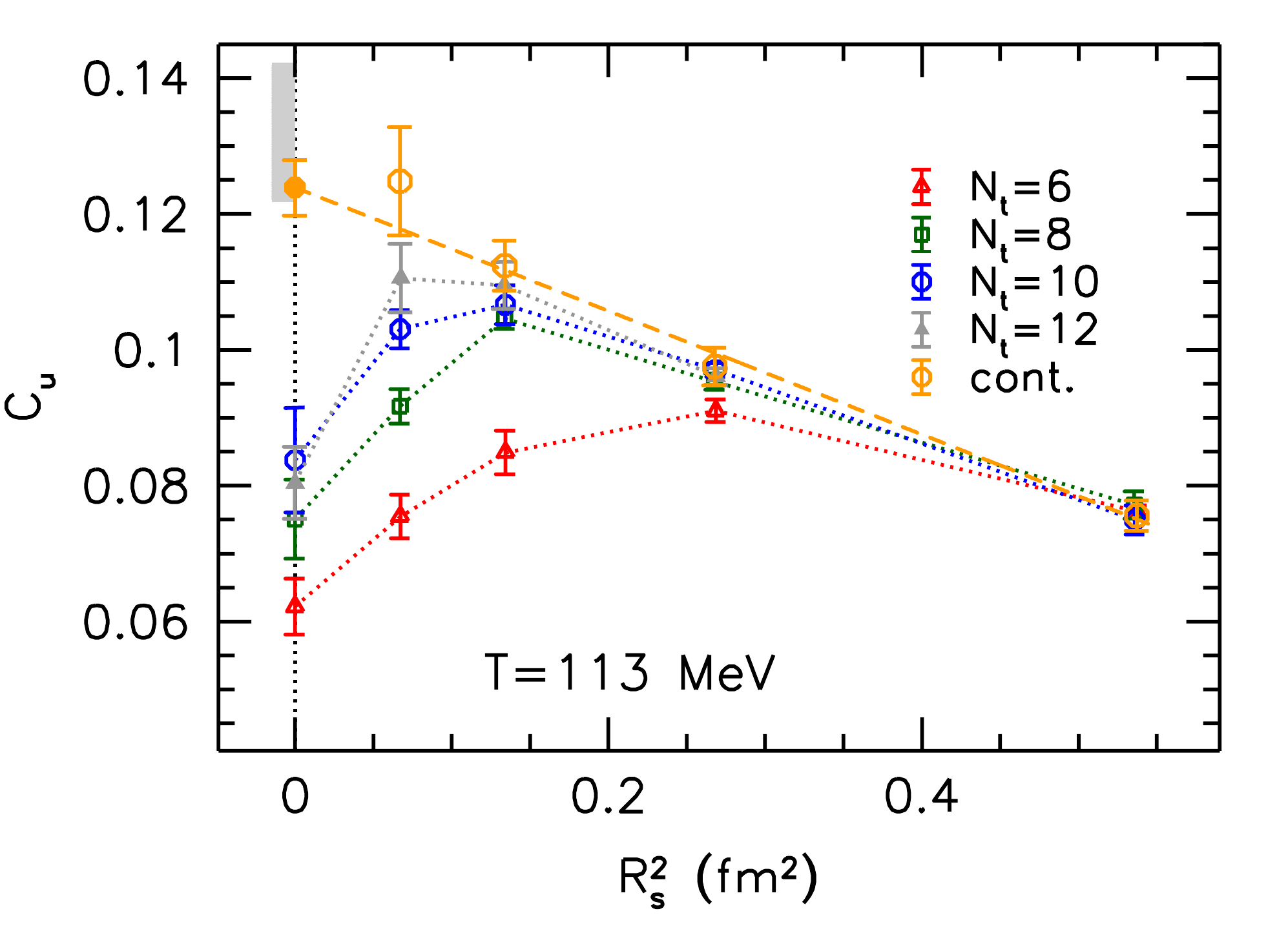} 
\includegraphics*[width=8.5cm]{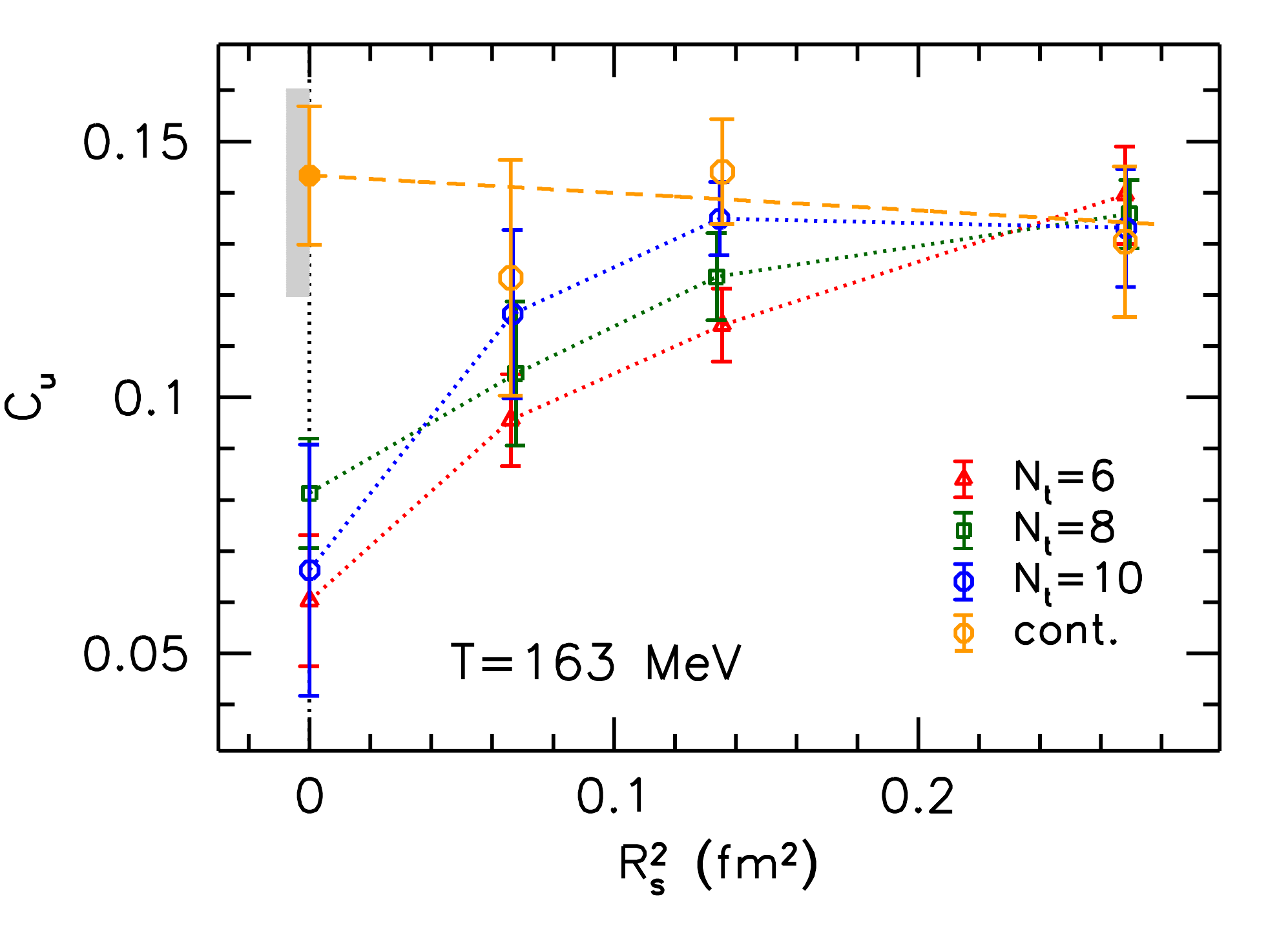} 
}
\caption{The coefficient $C_u$ as a function of the squared smearing range $R_s^2=8\,\ft^{\rm ph}$, introduced by the gradient flow, 
using four lattice spacings $N_t=6,8,10$ and $12$ at $T=113\textmd{ MeV}$ (left panel) and 
three lattice spacings $N_t=6,8$ and $10$ at 
$T=163\textmd{ MeV}$ (right panel). The dotted lines are to guide the eye. 
The continuum limit is performed 
at each $R_s$ (open yellow circles), followed by an extrapolation 
to $R_s=0$ (filled yellow circle). 
The error bars represent statistical errors and the gray region indicates the central value and total error 
of the final result 
obtained from a weighted histogram of many fits, see details in the text.
}
\label{fig:flowdep}
\end{figure}

We proceed with the renormalization, and investigate the effect of the gradient flow on the coefficient $C_f$. According to our expectations, $C_f$ is unphysical 
for $a\to0$ at vanishing flow time (vanishing smearing range), 
whereas for any nonzero $\ft^{\rm ph}\propto R_s^2$, it has a finite continuum limit. 
We demonstrate this in Fig.~\ref{fig:flowdep}, where $C_u(R_s^2)$ is 
shown for four lattice spacings at a fixed temperature $T=113\textmd{ MeV}$. 
While a power-type divergence is clearly absent from the $R_s=0$ data points, a logarithmic divergence cannot be excluded. 
At finite smearing ranges, we observe the convergence of the results to improve drastically 
-- at $R_s^2\approx 0.5 \textmd{ fm}^{2}$, the data points for all lattice spacings lie essentially on top of each other. Moreover, we also observe that the signal to noise ratio improves by up to an order of magnitude as the smearing range is increased beyond $1 \textmd{ fm}$.

For each $R_s>0$ dataset, we extrapolate the results to the continuum limit by a quadratic fit in the 
lattice spacing (motivated by the $\mathcal{O}(a^2)$ scaling properties of the action we use). 
For this extrapolation we use the three finest lattices and only include $N_t=6$ in the fit 
to estimate the systematic error. 
We find that the so obtained extrapolations are very well described by a linear 
function in $R_s^2$ (i.e., linear in the physical flow time $\ft^{\rm ph}$), which we 
use to extrapolate to $R_s=0$, see the left panel of Fig.~\ref{fig:flowdep} 
for the results for the up quark at $T=113 \textmd{ MeV}$.
We also consider a quadratic dependence on $R_s^2$, which we do not find to improve the fit qualities. 
Altogether, we take into account $2 \times 3\times 2 \times 2$ different fits (constant or quadratic fit in $eB$; including or excluding the point with the largest or the smallest $eB$; continuum extrapolation including or excluding $N_t=6$; linear or quadratic extrapolation in $R_s^2$ to $R_s=0$). 
The $a\to0$, $R_s\to0$ limits are used to build a weighted histogram, and the average value and 
systematic error is estimated -- following Refs.~\cite{Bali:1993ub,Durr:2008zz} -- 
by the mean and width of the obtained distribution, respectively.
(Fig.~\ref{fig:flowdep} shows one representative fit out of the many.)
The central values and the total (systematic and statistical) errors obtained 
from this procedure are given in Table~\ref{tab:res} and indicated by the gray regions at $R_s=0$ in Fig.~\ref{fig:flowdep}.

We perform a similar analysis in the deconfined phase, at $T=163\textmd{ MeV}$ using 
three ensembles with $N_t=6,8$ and $10$. 
The coefficient $\tau_f$ of the magnetic dipole moment quickly approaches zero as the temperature is increased, see 
Ref.~\cite{Bali:2012jv}. At the same time, the coefficient $\hat\tau_f$ of the 
topological charge density--electric dipole moment correlator 
is also found to drop, which lowers the signal-to-noise ratio in $C_f$. 
Moreover, we also observe that the continuum extrapolated 
data at $R_s>0$ show a much less pronounced dependence on $R_s^2$, as compared to the case 
at $T=113\textmd{ MeV}$, see the right panel of Fig.~\ref{fig:flowdep}. 
Motivated by this, in addition to the linear fits we also fit the data to a constant to extrapolate 
to $R_s=0$. The systematic error is again found by considering the width of the histogram 
built from results obtained by the various fit procedures. 

For the down quark -- again as a consequence of the $q_fB$-independence to leading order -- 
the results are within 
errors consistent with those obtained for the up quark. 
We find $C_s$ to be somewhat suppressed compared to the light quark 
coefficients, due to the larger mass of the strange quark. 
Our final results in the continuum limit at $R_s=0$, 
for the two temperatures under consideration, are shown in Tab.~\ref{tab:res}.
Note that the values for the two temperatures agree within errors for all flavors. 
Finally we remark that within our range of magnetic fields ($eB<0.5\textmd{ GeV}^2$), 
the behavior shown in the left panel of Fig.~\ref{fig:dipmoms} persists 
also at nonzero smearing ranges $R_s>0$ in the gradient flow, and the ratio of polarizations 
$\expvb{\qt(x)\cdot \Sigma_{zt}(x)}/\expvb{\Sigma_{xy}(x)}$ 
shows no significant dependence on $B$.

\begin{table}[ht!]
 \centering
 \begin{tabular}{c|c|c|c}
  $T$ & $C_u$ & $C_d$ & $C_s$ \\ \hline \hline
  $113 \textmd{ MeV}$ & 0.132(10) & 0.130(14) & 0.096(7) \\ \hline
  $163 \textmd{ MeV}$ & 0.14(2) & 0.12(3) & 0.09(2)\\
 \end{tabular}
 \caption{\label{tab:res}Continuum extrapolated results for the coefficient $C_f$ in the limit $R_s\to0$ at two values of the temperature.}
\end{table}

\begin{figure}[ht!]
\centering
\mbox{
\hspace*{-.5cm}
\includegraphics*[width=9.5cm]{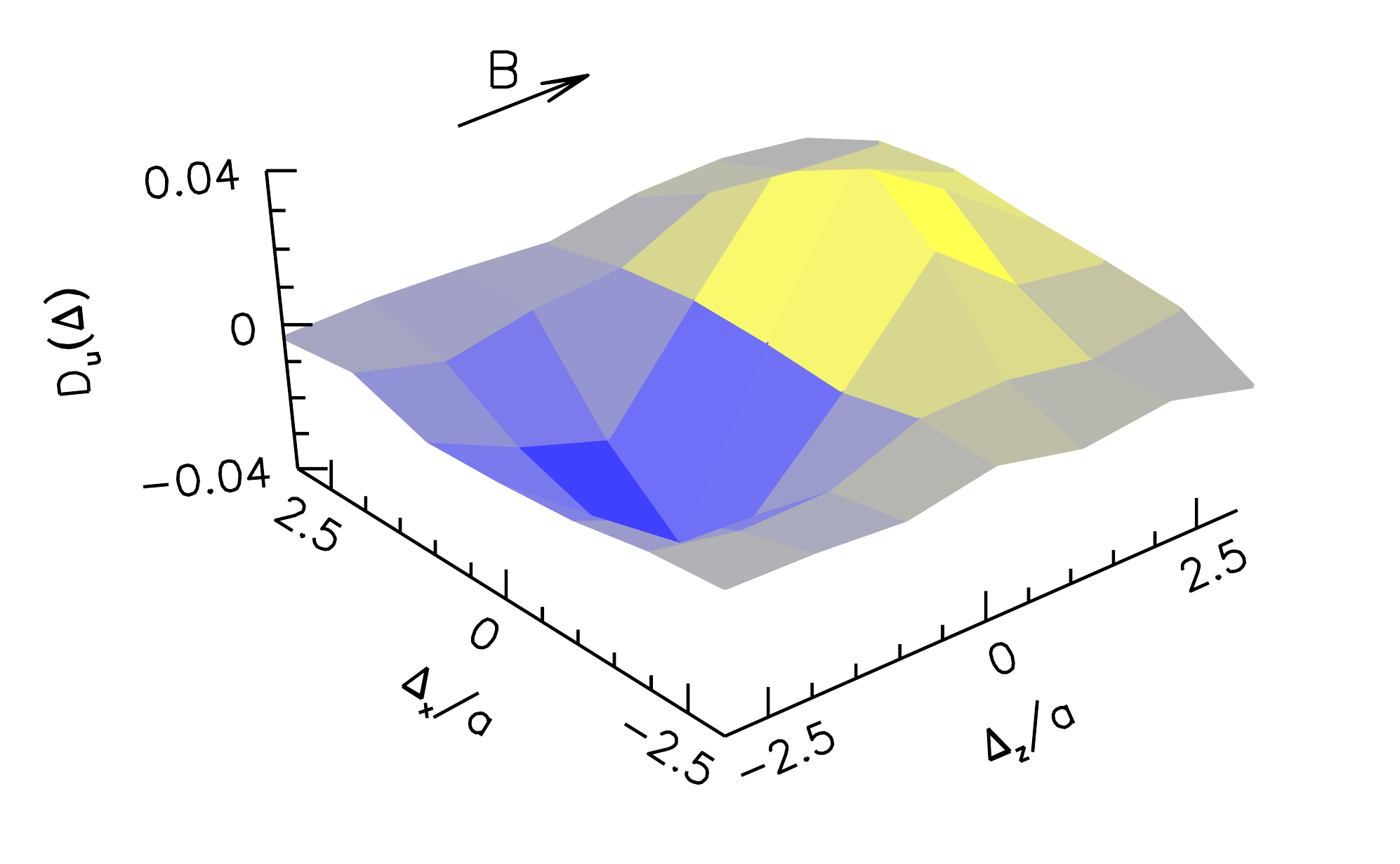}
\hspace*{-.7cm}
\includegraphics*[width=8.5cm]{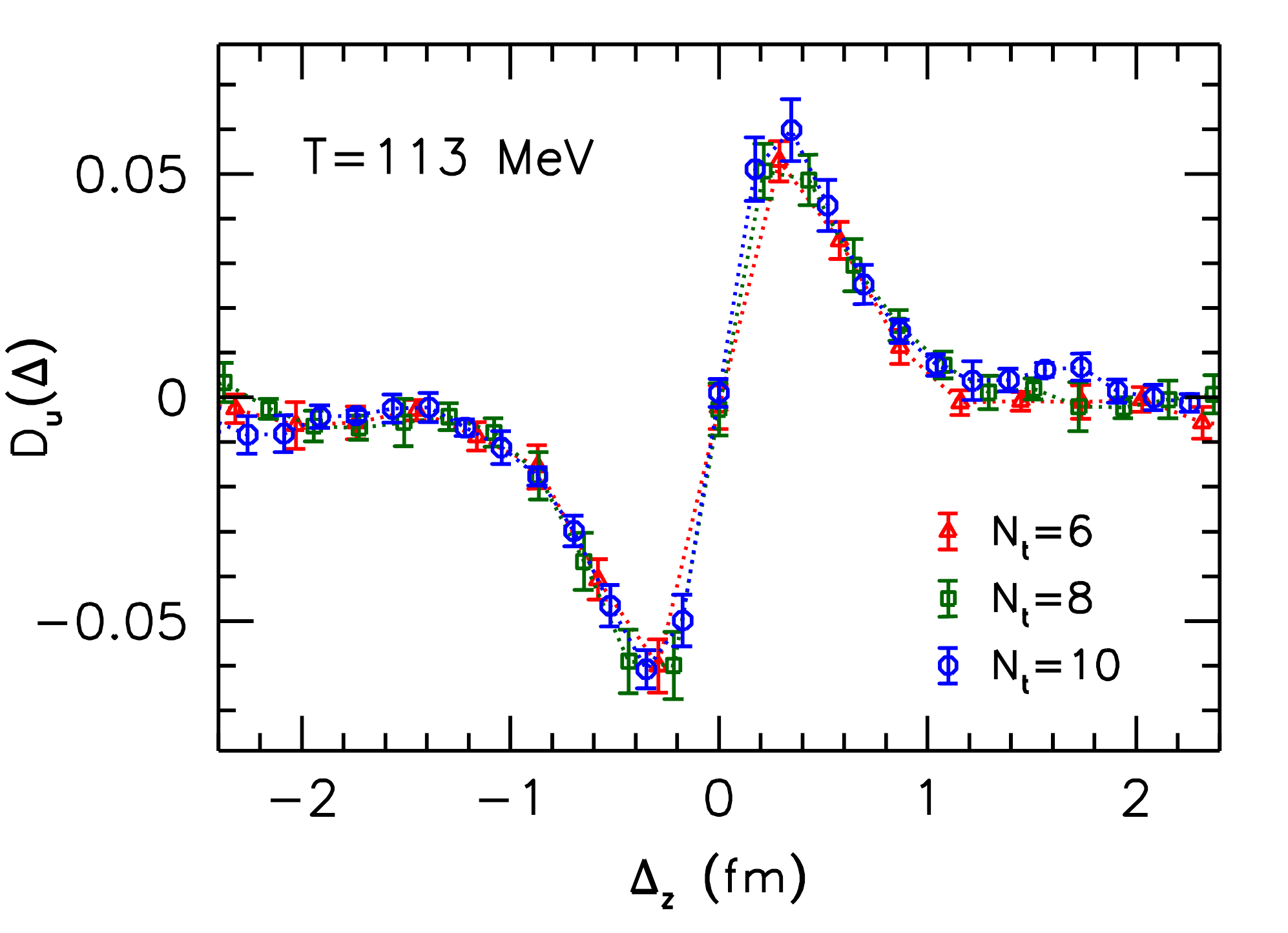}
}
\caption{
Left panel: extended dipole structure in the spatial electric charge density--topological charge density correlator 
in the $xz$ plane. The data was obtained on our $N_t=6$ ensemble at $T=113\textmd{ MeV}$ without any applied smearing. 
Right panel: spatial correlation along the $z$ direction for three lattice spacings at $T=113\textmd{ MeV}$ using a fixed smearing range $R_s^2\approx 0.27\textmd{ fm}^2$.
}
\label{fig:chargesep}
\end{figure}

Interpreting $\Sigma_{\mu\nu}^f$ as the electric dipole moment of 
the quark, it might seem that the induced polarization is point-like and is not related 
to spatial charge separation. 
However, due to the fluctuations in $\qt(x)$ and their interaction with dynamical sea quarks, 
the local electric dipole moment 
correlates with spatially extended dipole structures and, thus, with the spatial 
separation of the electric charge. 
To show that these extended structures exist, let us consider the electric current operator
\be
J^f_\nu(x)=\bar\psi_f\gamma_\nu \psi_f(x)
\ee
and compose the observable\footnote{
Note that in accordance with Lorentz-symmetry -- namely that $D_f$ should be antisymmetric in the two indices appearing in its definition -- the correlator involving $J_t^f$ along $\Delta_z$ 
equals minus the correlator involving $J_z^f$ along $\Delta_t$. 
}
\be
D_f(\Delta) = \frac{\displaystyle{}\expvB{\qt(x)\cdot J^f_t(x+\Delta)}}{\displaystyle{}\sqrt{\expvB{\qt^2(x)}} \expvB{\Sigma^f_{xy}(x)}},
\label{eq:Ddef}
\ee
where we employed the same normalization as in the definition~(\ref{eq:Cdef}) of $C_f$. 
The ratio $D_f(\Delta)$ represents the correlation between the topological charge density 
and the electric charge density at two distinct points separated by a four-vector $\Delta$. 
We remark that in our Euclidean setting, the correlator in the numerator of 
Eq.~(\ref{eq:Ddef}) is imaginary. 
Since the observable contains no dependence on the (imaginary) time, its analytic 
continuation simply amounts to a multiplication by $i$, giving a real observable in
Minkowski space-time\footnote{
To see this, note that 
the {\it density} of topological charge $\qt(x)$ receives a factor of $i$ via the continuation. 
Furthermore, the Minkowskian Dirac matrices are given by $\gamma_0^{\rm M} = \gamma_t$, $\gamma_i^{\rm M} = i\gamma_i$, such that the charge operator is the same in 
both space-times. Altogether, the observable $D_f$ gets multiplied by $i$. 
The same continuation for the observable $C_f$ gives no imaginary factor, since 
the spin operator in Minkowski space-time is defined as 
$\sigma_{\mu\nu}^{\rm M} = i/2\cdot [\gamma_\mu^{\rm M}, \gamma_\nu^{\rm M} ]$, such that 
$\sigma_{xy}^{\rm M} = \sigma_{xy}$ and $\sigma_{z0}^{\rm M} = -i \sigma_{zt}$. 
}. 
In the left panel of Fig.~\ref{fig:chargesep} we show this correlator for the up quark in the $xz$ plane. The figure reveals an excess of positive charge above ($\Delta_z>0$) the topological `source' 
and an excess of negative charge below ($\Delta_z<0$) it. Thus, we indeed observe an 
electric dipole 
structure aligned with the magnetic field. 

To show that this spatially separated electric charge is not a lattice artefact, in 
the right panel of Fig.~\ref{fig:chargesep} we plot $D_u(\Delta)$ for $\Delta = (0,0,\Delta_z,0)$. To approach the continuum 
limit in a well-defined manner, we again make use of the gradient flow and 
consider a nonzero smearing range. The 
results using three lattice spacings $N_t=6,8$ and $10$ lie almost perfectly on top of each other, 
showing small discretization errors and a fast scaling towards $a\to0$ -- similarly as we 
observed for $C_f$, compare Fig.~\ref{fig:flowdep}. 
For the strange quark (which exhibits a better signal to noise ratio) 
we also considered the dependence of the spatial integral 
$\int \d^3 \!\Delta \,D_s(\Delta)\,\Delta_z$ 
 -- corresponding to the electric dipole moment of the configuration -- on the 
 smearing range. 
The results indicate that this integral remains nonzero 
even in the limit $R_s\to0$.
Altogether, we conclude that the 
spatial separation of the electric charge remains a well-defined concept in the 
continuum limit.

\section{Comparison to a model}
\label{sec:comparison}

Let us now interpret our result for $C_f$ in a model and in the context of heavy-ion collisions. 
It is instructive to think of the quark-gluon plasma as depicted in the right 
panel of Fig.~\ref{fig:illustr}, with small independent domains containing gluon backgrounds 
of topological charge $\Qt$. In each domain an electric polarization $\Sigma_{zt}$ 
is induced by the magnetic field and by the local $\Qt$. 
This can be compared to the magnetic polarization $\Sigma_{xy}$, which 
is uniform in the whole volume. 
Let us further assume that the topological charge in each domain 
is created by \textit{constant} selfdual or antiselfdual non-Abelian fields 
of strength $G$, such that $\Qt\sim \pm G^2$. 
This is a generalization of the approach in Ref.~\cite{Basar:2011by} that allows to describe 
the case $B>G$ as well as $B<G$. Like in Ref.~\cite{Basar:2011by}, both polarizations in the 
local domain can be calculated analytically, when other gluonic interactions are neglected.
The calculation (for technical details, see App.~\ref{app:selfd}) simplifies 
tremendously in the lowest-Landau-level (LLL) approximation, which amounts to neglecting 
the quark masses compared to the field strengths. 
The magnetic field-dependence of the polarizations then reads\vspace*{-.3cm}
\be
\raisebox{-1.ex}{LLL, $m^2\ll G, |B-G|, |B+G|$:}\quad\quad
\begin{array}{ll}
 \Sigma_{zt}&= -\frac{1}{2\pi^2\,m} \cdot \si[\Qt]\, \begin{cases}
    G^2  & \text{for }G < B\\
    B G & \text{for }G > B
   \end{cases}\,,\\
 \Sigma_{xy}&= -\frac{1}{2\pi^2\,m}\cdot B G \,,
\end{array}
 \label{eq:parallel_fields_magn}
\ee 
where an identical proportionality factor has been neglected. 
This result is valid for one quark flavor (whose electric charge is set to unity) and gauge group SU(2) for spatially aligned Abelian and non-Abelian fields, and agrees with the calculation of Ref.~\cite{Basar:2011by}. (Note that Eq.~(\ref{eq:parallel_fields_magn}) does not hold for $G=0$ where the 
LLL approximation is invalid. In fact, in this limit $\Sigma_{zt}$ vanishes but $\Sigma_{xy}$ remains finite.)
In App.~\ref{app:selfd} we also discuss the case of non-aligned fields and gauge group $\mathrm{SU}(3)$ resulting in similar formulae. 

Based on our lattice results, we can make two important statements about this model. First, the ratio $\expv{ \integ \qt(x) \cdot \Sigma_{zt}(x)}/\expv{\integ\Sigma_{xy}(x)}$ is found to be $B$-independent for QCD with physical quark masses in the relevant range 
of magnetic fields,
cf. the left panel of Fig.~\ref{fig:dipmoms}. 
The equivalent of this quantity in one domain in the model treatment is $\Qt\cdot\Sigma_{zt}/\Sigma_{xy}$, which 
is independent of $B$ -- and, thus, reproduces the lattice findings -- only 
if the non-Abelian scale $G$ exceeds the external field $B$. 
Second, in this regime ($B<G$), we may compare the model prediction to the lattice 
results quantitatively. 
In order to compute the coefficient $C_f$ in the model, we need to assume 
a distribution of the topological charge among the local domains. 
A reasonable approximation is a Gaussian average\footnote{
A possible improvement of the model is to take into account 
correlations between the topological domains, similarly as in phenomenological 
instanton approaches, see, e.g., Ref.~\cite{Schafer:1996wv}. This might distort the 
Gaussian distribution of $Q_{\rm top}$.}
over $\Qt$.
In addition, we also consider an arbitrary angle $\vartheta$ between the non-Abelian 
and Abelian fields and integrate over $\vartheta$. This averaging over $\Qt$ and over $\vartheta$ is denoted by $\qdist{\ldots}$. 
Using Eq.~(\ref{eq:parallel_fields_magn}) for $B<G$ and its generalization to 
non-aligned fields, Eq.~(\ref{eq:generalization}) -- which we derived for $B\ll G$ -- we obtain
\begin{equation}
\textmd{LLL, }m^2\ll B \ll G:\quad\quad\quad
 C_f=\frac{\qdistb{ Q_{\text{top}}\cdot\Sigma_{zt} }}
 {\sqrt{\qdistb{ Q_{\text{top}}^2}}\,
 \qdistb{ \Sigma_{xy} }}
 =\frac{\qdistb{ |Q_{\text{top}}|^{3/2} \, \THET(\vartheta)}}
 {\sqrt{\qdistb{ Q_{\text{top}}^2 }}\,
 \qdistb{ |Q_{\text{top}}|^{1/2} \, \THET(\vartheta) }}=1.046,
 \label{eq:modelnumber}
\end{equation}
where $\THET(\vartheta)$ is a $\Qt$-independent factor describing the dependence of the 
polarizations on the angle. Its average over $\vartheta$ cancels in the ratio $C_f$. In fact, the above obtained number is independent of the width of the Gaussian distribution of $\Qt$ (due to the matching  powers of $\Qt$ in the 
numerator and the denominator). However, it differs from our lattice determination $C_f \sim 0.1$ by \textit{an order of magnitude}. Put differently, the strong interaction between quarks prevents their full polarization predicted by such a model.

The above 
comparison reveals that an effective description of QCD with magnetic fields has to take the strong interaction into account non-perturbatively and beyond the simple assumptions of this model.
In the same spirit one can question the lowest-Landau-level approximation used in the model 
setting. It corresponds to the idealized situation where the quark mass 
vanishes, and all quarks which are spin polarized by the magnetic field interact with 
the gluonic background and contribute to the electric polarization. 
However, heavier quarks are less sensitive to topology, and, accordingly, we expect 
the ratio $\Sigma_{zt}/\Sigma_{xy}$ to decrease as $m$ grows. 
This is consistent with our results $C_{u,d}>C_s$.

\section{Summary}
\label{sec:summary}

Using first principles lattice calculations, we have studied local CP violation in the QCD vacuum and its relation to the chiral magnetic effect, and determined the correlation coefficient between 
the electric polarization and the topological charge density, induced by an external magnetic 
field. We have considered $2+1$ flavor QCD 
with physical quark masses, and extrapolated the results to the continuum limit.
Our main result is a steady linear dependence of this correlation on $eB$ (without an indication of saturation) for magnetic fields $eB\lesssim 0.5 \textmd{ GeV}^2$, covering the maximal magnetic fields estimated to be present in heavy-ion collisions. 
The coefficient of proportionality -- after a normalization by the magnetic polarization, 
see Eq.~(\ref{eq:Cdef}) -- is obtained to be $C_f\sim 0.1$. The results for the three flavors $f=u,d,s$,  at two different temperatures are listed in Tab.~\ref{tab:res}.

We also estimated this coefficient 
using a model calculation employing nearly massless quarks, the lowest-Landau-level approximation 
and constant selfdual gluon backgrounds.
This model was found to overestimate $C_f$ by an order of magnitude. 
In other words, there is a substantial quantitative difference of the strength of local CP-violation for quasi-free quarks used in model approaches and fully interacting quarks in realistic physical situations.
Whether the electric current in the formulation of the CME with 
a chiral chemical potential~\cite{Fukushima:2008xe} 
is also subject to a similar suppression 
due to non-perturbative QCD effects
(first lattice results indicate a suppression by a factor of 3--4~\cite{Yamamoto:2011gk}), 
does not follow directly from our 
results. 
However, we take the results as a hint that effects 
due to local CP-violation in general contain similar suppression factors.

Let us finally add that we employed the staggered discretization of the 
QCD quark action in the lattice simulation, which in some topology-related aspects gives rise to large systematic/discretization  
errors. The topological susceptibility, $\expv{\Qt^2}/V$, for example, shows a rather slow 
scaling towards the continuum limit, see, e.g., Ref.~\cite{Bazavov:2010xr}. 
We find that for our particular observable, $C_f$, the continuum extrapolation 
is much flatter. This may have to do with the fact that $C_f$ is a local 
observable whereas the susceptibility is not. Nevertheless, it would be 
desirable to confirm our numerical findings with chiral fermions that have 
nicer topological properties.

\begin{acknowledgments}
This work was supported by the DFG (SFB/TRR 55, BR 2872/6-1), the EU (ITN STRONGnet 238353 and ERC No 208740) 
and the Alexander von Humboldt Foundation. 
The authors would like to thank Pavel Buividovich, Maxim Chernodub, Tigran Kalaydzhyan, Berndt M\"uller and K\'alm\'an Szab\'o for useful discussions.
\end{acknowledgments}

\appendix
\section{Details of the gradient flow}
\label{app:flow}

The smearing of the gluonic and fermionic fields is performed by evolving these fields in 
flow time $\ft$ 
($\ft^{\rm ph}=t\cdot a^2$ is the physical flow time). 
The evolution in flow time amounts to finding the solution of the flow equations for 
the gluonic links~\cite{Luscher:2010iy},
\be
\partial_\ft U^\ft = Z(U^\ft) U^\ft, \quad\quad U^{\ft=0} = U,
\ee
and for the quark fields~\cite{Luscher:2013cpa},
\be
\partial_\ft \psi_f^\ft = \Delta \psi_f^\ft, \quad\quad \psi_f^{\ft=0} = \psi_f,
\label{eq:fermflow}
\ee
and the corresponding equation for $\bar\psi_f^\ft$. 
Here, $Z(U^\ft)$ is the (algebra-valued) 
derivative of the plaquette action with respect to the link variable, 
and $\Delta$ is the lattice discretization of the Laplace operator (see below). 
The solution of the flow equations can be found by numerical integration, 
which is done using the third-order Runge-Kutta integrator described in 
Refs.~\cite{Luscher:2010iy,Luscher:2013cpa} (a stepsize of $\Delta\ft=0.02$ was found to be 
optimal here, see also Ref.~\cite{Borsanyi:2012zs}). 
Integrating the flow equations up to a fixed physical time $\ft^{\rm ph}=\ft \cdot a^2$ corresponds to a smearing of the fields 
over a range of $R_{s}=\sqrt{8\,\ft^{\rm ph}}$~\cite{Luscher:2010iy}. 

The definition of the quark condensate -- or, of the fermionic bilinears appearing in 
Eq.~(\ref{eq:Cdef}) -- at nonzero flow time requires the use of the adjoint flow for 
the noisy estimators $\eta_i$ of Eq.~(\ref{eq:measpol}) from flow time $\ft$ 
back to flow time $0$, see Ref.~\cite{Luscher:2013cpa}. 
For this, an optimal scheme for the storage of the evolved links 
$U^{\ft'}$ for $0\le\ft'<\ft$ is implemented. The evolution along the gradient 
flow is started from our original gauge action, thus with unsmeared links.
The stout smearing is then applied only for the measurement of the 
operators, see Eqs.~(\ref{eq:measFS})--(\ref{eq:measpol}).

We remark that there is a peculiar issue that arises if one applies the fermionic gradient flow 
for staggered quarks in a naive way. In the staggered fermionic discretization, the Dirac 
components of the quark field $\psi$ at site $x$ are distributed over vertices of the 
four-dimensional hypercube touching $x$. 
This distribution of the components is devised in a manner such that the staggered action 
becomes diagonal in Dirac space, 
and the only remnant of the original Dirac structure is through space-dependent real numbers, 
the so-called staggered phases. 
In particular, the mass term $\bar\psi\psi$ and the Dirac operator $\bar\psi \slashed{D} \psi$ 
are diagonal in Dirac space, therefore they can be represented in terms of the staggered 
quark fields $\chi$ in the same form, e.g. $\bar\chi\chi$. 
However, the naive discretization of the Laplace operator is not 
diagonal after the staggered transformation, giving no straightforward correspondence between the representation with the original fields and that with the staggered fields. 

To construct the Laplacian, let us define the forward and backward covariant difference operators,
\be
\nabla_\mu \psi(x)= U_\mu(x) \,u_\mu(x)\, \psi(x+\hat\mu) - \psi(x), \quad\quad\quad\quad
\nabla^\dagger_\mu \psi(x) = \psi(x) - U^\dagger_\mu(x-\hat\mu) \,u_\mu^*(x-\hat\mu) \,\psi(x-\hat\mu).
\label{eq:nablas2}
\ee
where $U_\mu\in\mathrm{SU}(3)$ are the gluonic links and $u_\mu\in\mathrm{U}(1)$ the phases 
corresponding to the magnetic field. 
The naive one-step discretization of the Laplace operator, 
$\Delta^{\rm naive}=\nabla^\dagger_\mu \nabla_\mu$ indeed becomes off-diagonal
as it mixes the tastes distributed over the hypercube in a non-trivial way.
One possibility to avoid this mixing of the tastes is to use the two-step discretization 
of the covariant differences,
\be
\begin{split}
\nabla^{(2)}_\mu \psi(x)&= \frac{U_\mu(x) \,u_\mu(x)  \,U_\mu(x+\hat\mu) \,u_\mu(x+\hat\mu) \,\psi(x+2\cdot\hat\mu) - \psi(x)}{2}, \\
\nabla^{(2)\dagger}_\mu \psi(x) &= \frac{\psi(x) - U^\dagger_\mu(x-\hat\mu) \, u_\mu^*(x-\hat\mu)  \,U^\dagger_\mu(x-2\cdot \hat\mu) \,
 u_\mu^*(x-2\cdot \hat\mu)
\,\psi(x-2\cdot\hat\mu)}{2}.
\end{split}
\label{eq:nablas}
\ee
to define the Laplacian 
$\Delta^{\rm diag}=\nabla^{(2)\dagger}_\mu \nabla^{(2)}_\mu$. This two-step discretization was used 
in the flow equation Eq.~(\ref{eq:fermflow}). 
The non-diagonal nature of $\Delta^{\rm naive}$ results in an explicit Lorentz-symmetry 
breaking of the evolved fermionic fields, even at $B=0$. This is indicated by asymmetric expectation values 
of the bilinear structures $\bar\psi\gamma_\mu \psi$. 
Using the two-step Laplacian $\Delta^{\rm diag}$, (the lattice discretized version of) 
Lorentz-symmetry is maintained, and $\expv{\bar\psi\gamma_\mu \psi}=0$ for all $\mu$. 

Finally we remark that we also attempted to use the square of the staggered Dirac operator in place 
of the Laplacian for the evolution of the fermionic fields in Eq.~(\ref{eq:fermflow}). 
The results obtained for the coefficient $C_f$ after the flow with $\slashed{D}^2$, however, 
showed an inferior scaling towards $a\to0$, as compared to the case with $\Delta^{\rm diag}$. Performing the 
extrapolation to the continuum limit was only feasible for the latter choice.

\section{Polarizations in topological backgrounds}
\label{app:selfd}
 
In order to evaluate $C_f$ in a topological background, we consider one quark flavor in constant commuting selfdual or antiselfdual non-Abelian fields, which exist in a finite Euclidean box with quantized fluxes~\cite{'tHooft:1981sz} (they can also be thought of as fields deep inside instantons or antiinstantons~\cite{Basar:2011by}) plus an Abelian magnetic field $B$. 
In these backgrounds, both the topological charge $\Qt$ and the polarizations 
$\Sigma_{zt}$ and $\Sigma_{xy}$ are constant in space. The quark mass is denoted by $m$ and 
the electric charge is set to unity for simplicity. Moreover, our notation is such that the QCD coupling does not enter the covariant derivative.
We follow two equivalent approaches to determine the electric and magnetic polarizations 
in this model setting. First we employ a spectral representation of the observables 
using Landau-levels. Second we write down the polarizations using the exact quark propagator 
in the specific background.

\subsection{Polarizations using the spectral representation}

Let us first consider the case where the non-Abelian field $G$ is 
(anti)parallel to the Abelian one $B$. Without loss of generality we can 
assume that $B$ points in the $z$ direction.
Taking $\mathrm{SU}(2)$ for the non-Abelian group, the $xy$ and $zt$ components 
of the total field strength $f$ read
\be
f_{xy}= \begin{pmatrix}
         B+G & 0 \\
	 0 & B-G \\ 
        \end{pmatrix}
, \quad\quad\quad
f_{zt}= \begin{pmatrix}
         \si[\Qt] G & 0 \\
	 0 & -\si[\Qt] G \\ 
        \end{pmatrix},
\label{eq:su2comps}
\ee
where we diagonalized the field strengths via a gauge transformation 
(for constant field strengths this is always possible). 
We also inserted the sign of the topological charge 
$\Qt = \pm G^2 /(2\pi^2)$ in the electric components to 
account for both the selfdual and the anti-selfdual cases.
Let us first discuss the upper color component and denote $b\equiv B+G$ and
$e\equiv \si[\Qt]G$.
The Dirac eigenvalues of this system are obtained through two independent 
Landau-level problems in the $(x,y)$- and $(z,t)$-planes (the arrows indicate 
the eigenvalues of the corresponding operators),
\begin{equation}
 -\slashed{D}^2=-D_\mu D_\mu+\frac{1}{2}\sigma_{\mu\nu}f_{\mu\nu}\,,\quad
 -D_\mu D_\mu \to |b|(2n_b+1)+|e|(2n_e+1)\,,\quad
\frac{1}{2}\sigma_{\mu\nu}f_{\mu\nu} \to s_b b+s_e e,
\end{equation}
with $n_b,n_e=0,1,\ldots$ and $s_b,s_e=\pm 1$.
The spin polarizations read~\cite{Bali:2012jv}
\begin{equation}
 \Sigma_{xy,zt}=m\, \text{tr}\,\frac{\sigma_{xy,zt}}{-\slashed{D}^2+m^2}
 = m \frac{|b| |e|}{4\pi^2} \sum_{
 \substack{n_b,s_b\\n_e,s_e}
 }
 \frac{s_{b,e}}{|b|(2n_b+1+s_b\si[b])+|e|(2n_e+1+s_e\si[e])+m^2},
\end{equation}
where $|b| |e|/(4\pi^2)$ is the degeneracy of all Landau-levels. The spin-dependence is such that 
only the corresponding lowest Landau-levels contribute: $\{n_b=0, s_b=-\si[b]\}$ for $\Sigma_{xy}$, 
whereas $\{n_e=0, s_e=-\si[e]\}$ for $\Sigma_{zt}$, 
cf.\ App.\ B in Ref.~\cite{Bali:2012jv}, giving
\begin{equation}\label{eq:pre_h}
 \Sigma_{zt}= -\frac{m\,e\,}{4\pi^2}\, h(b)\,,\qquad 
 \Sigma_{xy} = -\frac{m\, b\,}{4\pi^2}\, h(e)\,,\qquad
 h(f)\equiv |f|\sum_{n,s}\frac{1}{|f|(2n+1+s)+m^2}.
\end{equation}
Note that the polarizations change sign when their corresponding field strengths $e$ or $b$ are reversed, as they should. The sum in $h$ contains an $m$- and $|f|$-independent divergence,
\begin{equation}\label{eq:h}
 h(f)=-\frac{|f|}{m^2}+\sum_{k=0}^\infty\frac{1}{k+m^2/2|f|}
= -\frac{|f|}{m^2}+ \lim_{z\to1} \left[ \frac{1}{z-1} -\Psi^{(0)}(m^2/2|f|) + \mathcal{O}(z-1) \right],
\end{equation}
which we separated using zeta function regularization 
(here, $\Psi^{(0)}$ is the polygamma function of order $0$). 
The corresponding divergent contributions in the polarizations are linear in the field 
and can be absorbed into the renormalization of the electric charge\footnote{
Note that this is unnecessary for $\Sigma_{zt}$ since the divergence linear in $e=\si[\Qt]G$ 
cancels against the contribution of the second color sector, where $e=-\si[\Qt]G$, 
see \protect Eq.~(\ref{eq:su2comps}).
For $\Sigma_{xy}$ no such cancellation takes place since the magnetic field contains 
an Abelian component which is not traceless.
}.
After this subtraction, the leading term of the second contribution in Eq.~(\ref{eq:h}) 
in the limit $m^2\ll2|f|$ equals $2|f|/m^2$. It flips the sign of the first term in Eq.~(\ref{eq:h}) and, thus, for small masses $h=|f|/m^2$ indeed coincides with the lowest-Landau-level contribution, obtained by simply putting $n=0$, $s=-1$ in (\ref{eq:pre_h}). Hence, 
\begin{equation}\label{eq:one}
 (-4\pi^2\,m)\cdot\Sigma_{zt} =  e|b|\,,\qquad
 (-4\pi^2\,m)\cdot\Sigma_{xy} =  b|e|\,. 
\end{equation}

To calculate the full polarizations, we add the contributions of all 
color components in Eq.~(\ref{eq:su2comps}),\vspace*{-.2cm}
\be
\begin{split}
 \si[\Qt]\cdot(-4\pi^2\,m)\cdot\Sigma_{zt}&\,= G|G+B|+(-G)|-G+B|=2\begin{cases}
    G^2  & |G| < B\,, \\
    B|G| & |G| > B\, ,
   \end{cases}\\
 (-4\pi^2\,m)\cdot\Sigma_{xy}&\,= (G+B)|G|+(-G+B)|-G|=2B|G|,
\end{split}
 \label{eq:parallel_fields_two}
\ee
arriving at Eq.~(\ref{eq:parallel_fields_magn}) used in Sec.~\ref{sec:comparison}.
The first and third lines agree with Eqs.~(81) and~(82) of Ref.~\cite{Basar:2011by}, while the second line can also be obtained from the number of zero modes, Eq.~(47) of that reference.
Note that at $|G|=B$, where $\Sigma_{zt}$ would have a cusp, the lowest-Landau-level approximation breaks down in the color sector with field strength $-|G|+B$.

We now turn to the gauge group $\mathrm{SU}(3)$. One can again diagonalize the field strength, now it has two independent amplitudes in the fields $(G_1,G_2,-G_1-G_2)$ in three color sectors, and $|\Qt|\sim[G_1^2+G_2^2+(G_1+G_2)^2]$. This slightly complicates the calculations. For the simplest case of space-parallel fields in the lowest Landau-level  approximation one gets, in analogy to (\ref{eq:one})--(\ref{eq:parallel_fields_two})
\be
\begin{split}
 (-4\pi^2\,m)\cdot\Sigma_{zt}&\,= \si[\Qt]\big[G_1|G_1+B|+G_2|G_2+B|+(-G_1-G_2)|-G_1-G_2+B|\big],\\
 (-4\pi^2\,m)\cdot\Sigma_{xy}&\,=\qquad\qquad\:\: \big[(G_1+B)|G_1|+(G_2+B)|G_2|+(-G_1-G_2+B)|G_1+G_2|\big].
\end{split}
\ee
We have found that the ratio $\Sigma_{zt}/\Sigma_{xy}$ is $B$-independent and equals $\si[\Qt]$ when all three fields $|G_1|$, $|G_2|$ and $|G_1+G_2|$ are large compared to $B$, and that it is smaller and becomes $B$-dependent if one of them is not.

\subsection{Polarizations using the exact propagator}

We proceed by generalizing the above calculation and allow for an arbitrary polar angle $\vartheta$ between the non-Abelian and Abelian fields,
\be
\begin{split}
f_{xy} &= \begin{pmatrix}
         B+G\cos\vartheta & 0 \\
	 0 & B-G\cos\vartheta \\ 
        \end{pmatrix}
, \hspace*{2.48cm}
f_{xz}= \begin{pmatrix}
         G\sin\vartheta & 0 \\
	 0 & -G\sin\vartheta \\ 
        \end{pmatrix}
, \\
f_{zt} &= \begin{pmatrix}
         \si[\Qt]G\cos\vartheta & 0 \\
	 0 & -\si[\Qt]G\cos\vartheta \\ 
        \end{pmatrix}
,\quad
f_{yt}= \begin{pmatrix}
         -\si[\Qt]G\sin\vartheta & 0 \\
	 0 & \si[\Qt]G\sin\vartheta \\ 
        \end{pmatrix}.
   \\
\end{split}
\label{eq:su2comps_2}
\ee
This case should be equally relevant for estimating $C_f$ in realistic QCD 
configurations. We again considered the selfdual and the antiselfdual cases 
simultaneously by inserting $\si[\Qt]$ in the electric fields.

It is now advantageous to represent the polarizations (first we discuss a single color sector) by
\begin{equation}\label{eq:def_vevs}
 \Sigma_{xy,zt}=\text{tr}\, S(x,x)\sigma_{xy,zt},
\end{equation}
where $S$ is the Green's function of the Dirac operator in the presence of a constant Abelian field $f_{\mu\nu}$ and the trace contains a sum over spinor indices and an average over space-time. The latter is trivial since the field strength and also the polarizations are constant. For the Green's function we employ the proper time representation~\cite{Schwinger:1951nm},
\begin{equation}\label{eq:prop_time_int}
 S(x,x)=\frac{1}{16\pi^2}
\int_0^\infty \frac{dt}{t^2}\big[m+\mathcal{O}(\gamma_\mu)]
\exp\big[{-m^2 t}-L(-it)
-\sigma_{\mu\nu}f_{\mu\nu}t/2\big],
\end{equation}
where we have moved the integration contour from (just below) the real axis to the negative imaginary axis parameterizing the original integration variable as $s=-it\in[0,\infty)$. 
Here, $\mathcal{O}(\gamma_\mu)$ indicates terms that vanish under the Dirac trace in 
Eq.~(\ref{eq:def_vevs}), and the sign of the term containing $\sigma_{\mu\nu}$ is chosen such that 
it conforms to the definition~(\ref{eq:smndef}). Moreover, we introduced 
\begin{align}
\exp[-L(-it)]=\bigg[\det \frac{\sinh(-ift)}{-ift}\bigg]^{-1/2}\,,
\end{align}
viewing $f$ as an antisymmetric tensor in Lorentz indices (having purely imaginary eigenvalues).

Let us denote the invariants of $f$ 
(proportional to `action' and `topological charge' density) as
\begin{equation}
 u = \frac{f_{\mu\nu}^2}{4}\,,\qquad
 v = \frac{f_{\mu\nu}\widetilde{f}_{\mu\nu}}{4}\,, \qquad \widetilde{f}_{\mu\nu} = \frac{1}{2}\epsilon_{\mu\nu\alpha\beta}f_{\alpha\beta}\,,\qquad u>v.
\end{equation}
Then the eigenvalues are given by \cite{Schwinger:1951nm}
\begin{equation}
 is_1 \sqrt{u+s_2 \sqrt{u^2-v^2}}
 = i\,s_1\left( \sqrt{u+v} + s_2 \sqrt{u-v} \,\right) / \sqrt{2}
 \label{eq:evalues},
\end{equation}
with $s_1=\pm1$ and $s_2=\pm1$. The determinant of $f$ is simply $v^2$. The eigenvalues come in pairs with opposite signs, in accordance with the tracelessness of $f$, and the arguments of the square roots in them are all positive. Using this we obtain
\begin{equation}
 \exp[-L(-it)]=\frac{t^2|v|}{\sinh(\sqrt{u+\sqrt{u^2-v^2}}\,t)\sinh(\sqrt{u-\sqrt{u^2-v^2}}\,t)}\,.
\label{eq:B15}
\end{equation}
By explicit comparison we found that the other factor appearing in $S(x,x)\sigma_{\alpha\beta}$ can be represented as 
\begin{equation}
 \text{tr}\,\big[e^{-\sigma_{\mu\nu}f_{\mu\nu}t/2}\sigma_{\alpha\beta}\big]
 =-\sqrt{2}
  \bigg[\frac{\sinh(\sqrt{2(u-v)}\,t)}{\sqrt{u-v}}(f-\widetilde{f})_{\alpha\beta}
      +\frac{\sinh(\sqrt{2(u+v)}\,t)}{\sqrt{u+v}}(f+\widetilde{f})_{\alpha\beta}\bigg].
\end{equation}
For our situation these quantities read
\begin{align}
 u=&\,G^2+BG\cos\vartheta+\frac{B^2}{2}\,,\qquad 
 v=\si[\Qt]\,G\,(G+B\cos\vartheta)\,,\\
 2(u+\si[\Qt] v)=&\,4G^2+4GB\cos\vartheta+B^2\,,\qquad
 2(u-\si[\Qt] v)=B^2,\label{eq:second_line}
\end{align}
and in terms of
\begin{equation}\label{eq:w}
  w=\sqrt{2(u+\si[\Qt] v)}=\sqrt{4G^2+4GB\cos\vartheta+B^2}\,,
  \qquad z=\frac{2G\cos\vartheta+B}{w},
\end{equation}
the projections become
\be
\begin{split}
 \text{tr}\,\big[e^{-\sigma_{\mu\nu}f_{\mu\nu}t/2}\sigma_{xy}\big]
   &=\,-2\,\big[ z\sinh(wt)+\sinh(Bt)\big] , \\ 
 \text{tr}\,\big[e^{-\sigma_{\mu\nu}f_{\mu\nu}t/2}\sigma_{zt}\big]
   &=\,-2\,\big[z \sinh(wt)-\sinh(Bt)\big]\,\si[\Qt] . 
\end{split}
\label{eq:B20}
\ee
The second line confirms that $\Sigma_{zt}$ changes sign if the topological charge does so.

Plugging all this into the proper time integral (\ref{eq:prop_time_int}) shows that 
the integral diverges as $t\to 0$. 
This is the same divergence that we encountered in Eq.~(\ref{eq:h}). 
Here we eliminate it by dividing the observable by $m$ and differentiating it with respect to $m^2$, cf. Eq.~(\ref{eq:pre_h}). This indeed renders the integral finite and also reveals that the divergence is 
independent of $m$ and of the fields. 
Setting $\vartheta=0$ (and consequently $z=\pm 1$ etc.) reproduces the finite part of Eq.~(\ref{eq:h}). 

For $\vartheta\neq0$, two hyperbolic sine functions are left in the denominator of Eq.~(\ref{eq:B15}), such that the proper time integral cannot be performed easily. Since we argued that the region $|G|>B$ (i.e.\ $|G|>B\cos\vartheta$ for all $\vartheta$) is the relevant one for comparison with the lattice data, we now specialize to this case. 
Then $w > B$ and the proper time integral reads\footnote{
In the general case one has to use $|w-B|$ instead of $w-B$ in the denominator of \protect Eq.~(\ref{eq:B21}).}
\begin{align}
 \frac{\partial\Sigma_{xy}/m}{\partial m^2}
  =\frac{|v|}{8\pi^2} \,\int_0^\infty \!dt\, \exp\big({-m^2 t})\,t\,
  \frac{z\sinh(wt)+\sinh(Bt)}
       {\sinh\big(\frac{w+B}{2}t\big)\sinh\big(\frac{w-B}{2}t\big)}.
\label{eq:B21}
\end{align}
Similarly as in Eq~(\ref{eq:one}), we now resort to the approximation $m^2\ll |G|,B$. Moreover, 
in order to simplify the integral, we also assume $B\ll |G|$. 
Expanding the fraction in $B/|G|$ and $m^2/|G|$, we can perform the $t$-integral to arrive at
\be
 \frac{\partial\Sigma_{xy}/m}{\partial m^2} 
=\frac{|v|}{8\pi^2} \cdot \left[
\frac{2z}{m^4}+\mathcal{O}\Big(\frac{1}{G^2}\Big)
+\mathcal{O}\Big(\frac{B}{G^3},\frac{Bz}{G^3},\frac{m^2}{G^3}\Big) \right].
\ee
Notice that the term $\sinh(Bt)$ in Eq.~(\ref{eq:B21}) does not contribute at this order, 
which, using Eq.~(\ref{eq:B20}), implies that $\Sigma_{xy}=\Sigma_{zt}\,\si[\Qt]$. 
Using the expansion $z=\si[G](\cos\vartheta+B/(2G)\cdot\sin^2\vartheta+\mathcal{O}(1/G^2))$ and $|v|=G(G+B\cos\vartheta)$ gives
\begin{align}
\frac{\partial\Sigma_{xy}/m}{\partial m^2}
  =\frac{1}{4\pi^2}\frac{|G|}{m^4}\big(G \cos\vartheta+B\cdot\THET(\vartheta)\big)
  +\mathcal{O}(G^0)+\mathcal{O}\Big(\frac{B}{G},\frac{m^2}{G}\Big),\qquad
\THET(\vartheta) = \cos^2\vartheta+\frac{\sin^2\vartheta}{2}.
\end{align}

Here the leading term $\sim|G|G$ vanishes upon adding the second color sector of $\mathrm{SU}(2)$, which amounts to the same expression with $G\to -G$, cf. Eq.~(\ref{eq:parallel_fields_two}).
Adding the contributions of both sectors and integrating in $m^2$ we finally get
\be
(-4\pi^2\,m)\cdot\Sigma_{xy} = 2B|G|\cdot \THET(\vartheta), \qquad
(-4\pi^2\,m)\cdot\Sigma_{zt} = \si[\Qt]\cdot 2B|G|\cdot \THET(\vartheta),
\label{eq:generalization}
\ee
which, at $\vartheta=0$, reproduces Eq.~(\ref{eq:parallel_fields_two}) for the case $B<G$.
This expression was inserted in Eq.~(\ref{eq:modelnumber}). Note that the average over 
the polar angle factorizes and gives
\be
\frac{1}{2}\int_0^\pi\d\vartheta \sin \vartheta \, \THET(\vartheta) = \frac{2}{3}\,.
\ee

\bibliographystyle{jhep}
\bibliography{cme}

\end{document}